# *Onesweep*: A Faster Least Significant Digit Radix Sort for GPUs

Andy Adinets
NVIDIA Corporation
aadinets@nvidia.com

Duane Merrill
NVIDIA Corporation
dumerrill@nvidia.com

## ABSTRACT

We present *Onesweep*, a least-significant digit (LSD) radix sorting algorithm for large GPU sorting problems residing in global memory. Our parallel algorithm employs a method of single-pass prefix sum that only requires ~$2n$ global read/write operations for each digit-binning iteration. This exhibits a significant reduction in last-level memory traffic versus contemporary GPU radix sorting implementations, where each iteration of digit binning requires two passes through the dataset totaling ~$3n$ global memory operations.

On the NVIDIA A100 GPU, our approach achieves 29.4 GKey/s when sorting 256M random 32-bit keys. Compared to CUB, the current state-of-the-art GPU LSD radix sort, our approach provides a speedup of ~1.5x. For 32-bit keys with varied distributions, our approach provides more consistent performance compared to HRS, the current state-of-the-art GPU MSD radix sort, and outperforms it in almost all cases.

## 1 INTRODUCTION

Sorting is germane to many problems in computer science. As a preprocessing step, sorting facilitates a diversity of search and query problems [4,10]. Sorting also plays an important role in the construction and manipulation of data structures, especially those modeling relationships in sparse systems [3,11,16,21]. Furthermore, the discretionary and/or partial sorting of tasks and data is often undertaken as a performance enhancement, e.g., to improve spatial/temporal locality, to reduce conflicts and contention, to increase communication efficiency, etc.

In the last decade, design trends in computer architecture have embraced wider, fine-grained parallelism to deliver greater performance while maintaining energy efficiency. Modern GPU processors exemplify this trend, with each processor provisioning hundreds of thousands of data parallel hardware thread contexts. In the landscape of GPU sorting methods, radix sorting algorithms have consistently demonstrated the highest levels of performance when sorting numeric key types [1,17]. The work presented in this paper significantly improves the state of the art in parallel, least-significant-digit (LSD) GPU radix sorting, specifically for problems residing in global memory or last-level cache, i.e., data sets in the range of 128 KiB to 80 GiB.

As a lexicographic sorting method, radix sorting relies upon a representation of key data comprising an ordered sequence of digits. For each digit-place of $d$ bits within a $k$-bit key, a radix-$r$ sorting method will partition the keys into $r = 2^d$ bins based upon their digits at that digit-place. Digit binning is performed iteratively across all $p = \lceil k/d \rceil$ digit places, proceeding either from the most-significant to least-significant digit place (MSD sorting), or vice versa (LSD sorting). Consequently, the overall work complexity for either variant is $O(kn)$, a linear function of problem size.

MSD and LSD radix sorting have their individual strengths and challenges. The attractiveness of MSD radix sorting is the ability to exploit the hierarchical memories of processors such as GPUs. Digit bins are recursively partitioned into sub-bins at each iteration, shrinking in size until either empty, comprising a singleton key, or the last digit place has been processed. When a sub-problem falls below a certain size threshold, the remaining digit places can be processed entirely in-core using faster local memory. For dataset sizes and key distributions that quickly devolve into sub-problems small enough for "local-finishing", recent work has indicated a general performance advantage for MSD variants in global-memory sorting scenarios [9,18].

With LSD radix sorting, however, each digit-binning iteration repartitions the entire dataset into $r$ bins residing in the original memory domain. Although there is no equivalent opportunity for local-finishing, LSD methods have historically provided two advantages relative to MSD: (1) minimal bookkeeping overhead and complexity from only having to manage one binning instance at a time; and (2) a near-uniform performance response, regardless of key distribution. Additionally, LSD variants are necessarily *stable*, i.e., the relative order of same-valued keys in the sorted output is the same as the unsorted input. Stability is a common requirement of many sorting scenarios.

Our *Onesweep* algorithm for parallel LSD radix sorting introduces a third performance benefit: the ability to perform each digit binning iteration in a single pass through the dataset using only ~$2n$ global read/write operations. In contrast, contemporary parallel LSD and MSD implementations perform two full passes through the data for each digit-binning iteration: (1) an "upsweep" pass to compute per-block digit histograms, and (2) a "downsweep" pass to relocate keys. Together, these passes require ~3*n* global memory operations. [15,18].

Our method of single-pass digit binning derives from the *chained scan* parallelization of single-pass prefix sum presented by Merrill *et. al* [13]. We compute per-block bin offsets in the digit binning pass itself by propagating prefix totals in a manner that hides the latency of inter-block communication. This obviates the need to compute per-block histograms in separate pass. Only the prefix sum of global digit counts is needed prior to binning. In *Onesweep*, we compute the global digit counts for *all* digit places within a single histogram pass prior to any digit-binning iterations. Assuming memory-limited performance tuning, the elision of *p*-1 histogram passes over the course of an entire LSD radix sort provides a theoretical speedup ceiling of 1.5x vs. prior LSD radix sorting designs.

LSD radix sorting permits this upfront consolidation of histogram work because the binning scope is always global, i.e., the digit populations within each digit place are iteration-invariant. A similar optimization cannot be made for MSD methods because the digit population of each subproblem is dependent upon that of its predecessor. Consequently, *Onesweep* will also afford a speedup ceiling of 1.5x versus MSD methods for sorting problems that do not quickly converge to "local-finish" work.

We remark that Duvanenko recently made an independent, similar observation regarding the upfront consolidation of digit histograms for sequential implementations of LSD radix sorting [6]. Critically, it is our combination of upfront histogram computation followed by an augmented structure of *chained scan* digit binning that enables high performance, single pass partitioning on a parallel machine.

In this paper, we make the following contributions:

- We propose *Onesweep*, a parallel LSD radix sorting algorithm based on single-pass prefix sum that only requires ~2*n* memory operations per global-binning iteration.

- We measure the performance of our approach and evaluate it vs. the state-of-the-art in GPU LSD and MSD radix sorting. We achieve the sorting speed of 29.4 GKey/s for uint32 keys on A100.

- Our implementation has been integrated into the CUB open-source library (v 1.10) [12] and is freely available.

   Compared to the prior implementation in CUB, the previous state-of-the-art GPU LSD radix sort, we achieve a speedup of 1.4x-1.6x with different key and value types and different key distributions.

- Our digit-binning implementations efficiently support the simultaneous, concurrent computation of $r = 256$ prefix sums. This affords an 8-bit digit place, allowing us to reduce the overall number of binning iterations relative to the prior CUB LSD implementation (e.g., from five passes down to four for 32-bit keys).

- Compared to HRS [18], the state-of-the-art GPU MSD radix sort, we provide more consistent performance. For varied key distributions, we outperform HRS in almost all cases.

## 2 BACKGROUND

*GPU programming model and terminology.* In GPU algorithm design, the most efficient implementations conform to the hierarchical, bulk-synchronous nature of the machine's programming model [7]. A sorting computation will proceed as a sequence of *kernel* invocations, each a short program executed by a hierarchical *grid* of threads. Individual threads are grouped into *thread blocks* which are themselves grouped into grids. The thread block abstraction virtualizes the hardware's streaming multiprocessor cores (*SM*s). Threads within the same block can cooperate through fast in-core *shared memory* and local barrier synchronization. Data flow among SMs is typically bulk synchronous, i.e., the overall computation is orchestrated as a sequence of kernel invocations where the threads of each invocation are presented with a consistent view of any *global memory* updates from the previous ones.

*GPU sorting*. The recent GPU sorting surveys by Arkhipov *et al*. [1] and Singh *et al*. [17] provide excellent historical and comparative expositions spanning a large diversity of methods and scenarios. For global sorting problems, the advantage of radix sorting methods relative to their comparison-based counterparts (e.g., mergesort, quicksort, bitonic sort, etc.) stems from their ability to trade local computation for last-level memory bandwidth. Specifically, increasing the width *d* of the radix digit has the effect of decreasing the total number of last-level memory accesses by reducing the overall number of binning iterations. This tradeoff presents an "uphill" proposition, however: a linear increase in *d* corresponds to an exponential increase in dynamic instruction counts and local storage requirements, which scale with the radix *r*. As an example of per-architecture tuning over the last seven generations of GPU microarchitecture, the LSD sorting radix in NVIDIA's

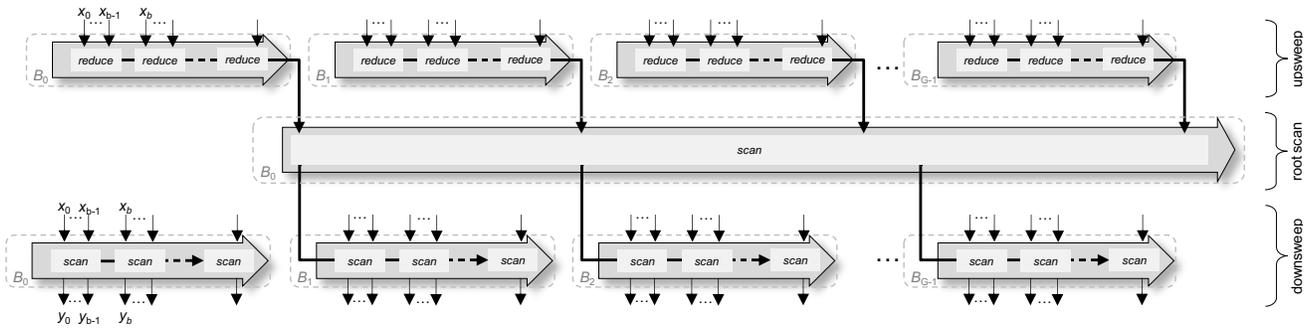

**Figure 1.** Three-kernel *reduce-then-scan* parallelization among *G* thread blocks (~3*n* global data movement) [12]

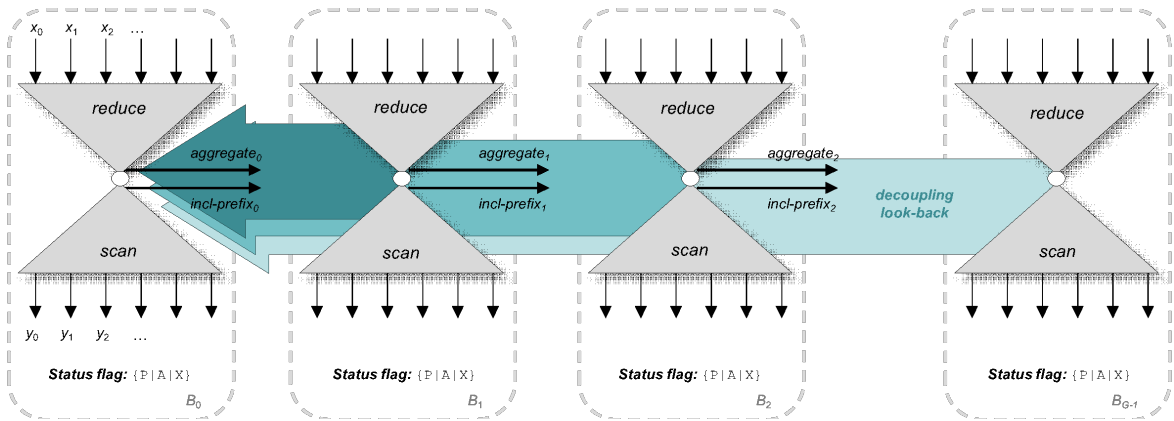

**Figure 2.** Single-pass *adaptive look-back* prefix scan among *G* thread blocks (~2*n* global data movement) [12]

CUB library has increased from $r = 16$ (eight digit-binning iterations) to $r = 128$ (five digit-binning iterations) for large 32-bit sorting problems [12].

In contrast, comparison-based sorting methods have super-linear $O(n\log n)$ asymptotic work complexity. Furthermore, practicable strategies for reducing the number passes through global memory have been elusive. For example, state-of-the-art GPU merge sorting implementations perform binary (radix-2) merging [5,8]. As such, a merge sort of 16M 32-bit keys will require ~10 passes through global memory (i.e., one pass to bootstrap 32K sublists followed by nine binary merging passes), whereas our *Onesweep* LSD radix sort only requires five (i.e., one histogram pass followed by four binning iterations).

*Prefix sum*. An *exclusive prefix sum* across a list of numbers produces a corresponding list in which the $i^{th}$ output is the sum of the first $i$-1 inputs. It is a useful construct for allocation-like behavior in parallel settings (partitioning, binning, compaction, queuing, etc.), where the offset for the $i^{th}$ thread to write its output is the sum of the number of output items being produced by the prior $i$-1 threads.

Prefix sum plays two roles in digit binning. The first is determining the starting location of each bin in the output buffer such that no space is wasted. For example, an exclusive prefix sum across the radix $r = 8$ histogram `<8,6,7,5,3,0,9,2>` of digit counts produces the corresponding list of offsets `<0,8,14,21,26,29,29,38>` indicating the locations of the `0`s, `1`s, `2`s, etc. bins in the output buffer.

Prefix sum is also used to compute the bin-relative offsets for scattering keys into their destination bins. Specifically, a prefix sum across a list of binary flags indicating which keys contain a given digit will produce a corresponding list of scatter offsets within that digit's output bin. Given the list of keys `<17,8,24,5>`, for example, the flag list `<0,1,1,0>` corresponds to the presence of a `0`s digit in the least-significant 3-bit digit-place. A prefix sum of these flags produces the scatter offsets `<0,0,1,2>` for relocating the flagged keys relative to the start of the `0`s bin. Thus, each digit-binning iteration entails computing $r$ prefix flag sums, one for each bin.

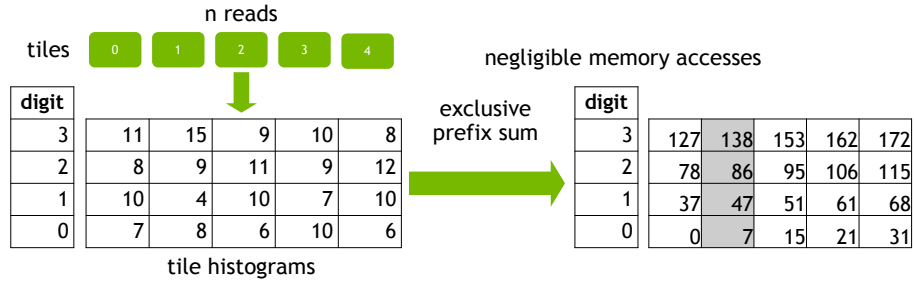

(a) The *upsweep* produces per-block digit histograms. A small *r×g* element exclusive prefix sum of those concatenated counters establishes sets of per-block bin offsets.

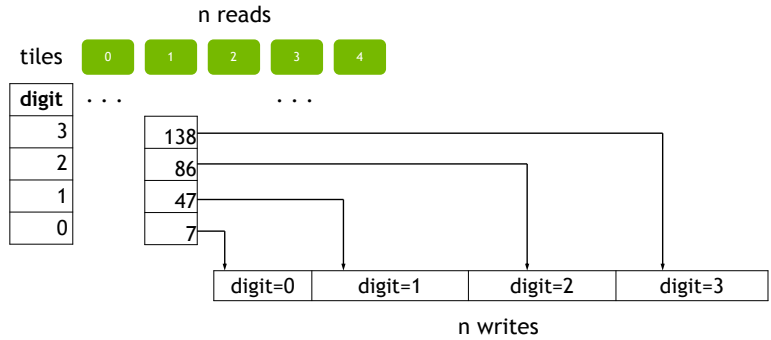

(b) The *downsweep* uses these offsets to indicate where each block can begin scattering its keys into the different digit bins. (E.g., tile 1 will begin writing its **1**s digits to offset 47.)

**Figure 3. An example of *reduce-then-scan* bookkeeping for a digit binning iteration of a *d*=2 bit digit place into *r*=4 bins, with input comprising *g*=5 tiles of 36 elements each.**

Historically, GPU prefix sum implementations have embodied one of two strategies at the global level: *reduce-then-scan* [14] or *chained scan* [13,20]. The former requires three kernels and two full passes through the data for ~3*n* total memory operations. As illustrated in Figure 1, each block in the first kernel serially reduces the tiles of its partition. Then the second kernel performs a prefix sum across the small list of per-block aggregates. Finally, each block in the third kernel computes a serial prefix sum across the tiles of its partition, seeded by its per-block prefix computed by the second kernel.

The *chained scan* approach is implemented by a single kernel where each block is assigned a tile of input, and a serial chain of prefix dependences exists between them. Instead of being computed by a previous kernel, the running prefix total propagates directly from block to block. As illustrated in Figure 2, the latency of this propagation can be hidden via a "decoupled lookback" strategy in which each block generally free to progressively consume the per-tile totals recorded by its predecessors until it discovers one that has also recorded the global inclusive prefix up to and including that tile. The method trades some redundant computation within each block for the ability to avoid prolonged waiting on its immediate predecessor. It must only wait on the processor to produce its local tile-aggregate (versus the entire running total), and all blocks are expected to finish producing their aggregates at roughly the same time. This latency hiding allows the computation to proceed with near-maximal bandwidth utilization.

We also note that atomic read-modify-write operations can provide similar utility as prefix sum with less overhead, especially on architectures where they are implemented at the memory-level. Atomics have several drawbacks, however: (1) their update-order is non-deterministic, preventing them from being used in stable LSD digit binning outside of simple digit-counting tasks; (2) performance can be capricious in sorting problems having low digit diversity due to contended accesses to the same counters; and (3) their hardware support may be non-existent or have insufficient throughput for use at scale.

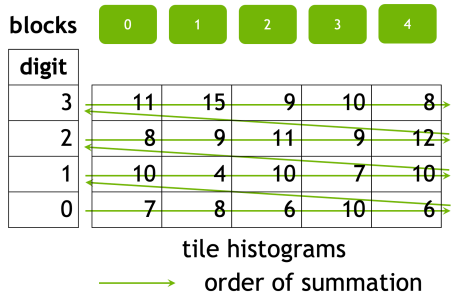

**Figure 4.** The prefix summation order of per-block digit counts for *r*=4 digits and *g*=5 blocks.

*Reduce-then-scan digit binning.* Contemporary designs for parallel LSD (and MSD) digit-binning employ a *reduce-then-scan* architecture that has been extended to compute *r* concurrent prefix sums. As illustrated in Figure 3, each digit-binning iteration performs the following three steps:

1. *Upsweep.* In a full pass through the key data, thread blocks produce histograms of digit counts at the current digit place, one histogram per block. The global memory workload is dominated by the *n* reads needed for key inspection and digit decoding.

2. *Prefix sum of block counts.* Produces a set of digit-bin offsets for each block by computing a prefix sum across the concatenated histograms of per-block digit counts. The overhead of this step is typically insignificant relative to the other two, as the *r×g* product of digit counts and thread blocks is typically ≪ *n*.

3. *Downsweep.* In a second full pass through the data, thread blocks perform a stable partitioning of elements into bins. Keys are loaded and decoded again to produce digit flags. Their scatter offsets are computed via prefix sums of flag data, seeded with the per-block offsets computed during the prefix sum phase. The global memory workload is dominated by the gather-scatter cost of *n* reads and *n* writes.

The total workload comprises ~3*n* global memory operations. However, each binning iteration would ideally incur only *n* reads and *n* writes. Unfortunately, a direct appropriation of *chained scan* cannot consolidate all three steps in a single pass. At least one separate pass through the data is needed before it can be partitioned. This is due to the forward dependence of each bin's starting offset on the cumulative totals of lesser digits, as shown in Figure 4. For example, the offset for scattering keys with 1s digits requires knowing the global count of all keys with 0s digits.

## 3 ONESWEEP LSD RADIX SORTING

*High-level description.* In adapting *chained scan* for LSD digit binning, we observe the output offset at which each thread block begins writing its elements for a given digit is a sum of:

1. *Global bin offset*: the starting index into the global output at which all elements with a given digit are to be binned.

2. *Bin-relative block offset*: the index into a given bin at which a specific thread block will begin placing elements having the corresponding digit.

For a given digit place, the global bin components can be obtained by first computing a global histogram of digit counts and then applying an exclusive prefix sum across it. Additionally, we note the histogram of digit counts for each digit place is independent of key order. Therefore, we can compute the histograms for all binning iterations in a single upfront pass. We compute the global components for every bin in every digit place by applying an exclusive prefix sum across each histogram.

Furthermore, a block's bin-relative component for a given digit is independent of the keys being sent to other digits. It depends only on the key-counts of previous blocks for that digit. Therefore, the bin-relative components for each block can be produced during the single-pass execution of an augmented *chained scan* that fuses *r* concurrent global prefix sums, one per radix bin.

Leveraging these observations, the overall *Onesweep* sorting procedure consists of the following three phases, each of which is implemented as a separate GPU kernel:

1. *Histograms of global digit counts.* Computes digit histograms for all digit places in a single pass.

2. *Prefix sums of global digit counts.* Computes the global bin offset for every digit in each digit place.

3. *p = ⌈k/d⌉ iterations of chained scan digit binning.* For each iteration, each thread block reads its tile of elements, decodes key digits, participates in a *chained scan* of block-wide digit counts, and scatters its elements into their global output bins.

Consequently, *Onesweep* performs a total of ~(2*p*+1)*n* memory operations across the entire sort. In contrast, *reduce-then-scan* requires ~3*pn* memory operations for sorting. Thus, *Onesweep* reduces the number of memory accesses by a factor of 3*p*/(2*p*+1). As a typical example, a *Onesweep* sort of 32-bit keys using 8-bit digit places will provide a 1.33x reduction in memory accesses.

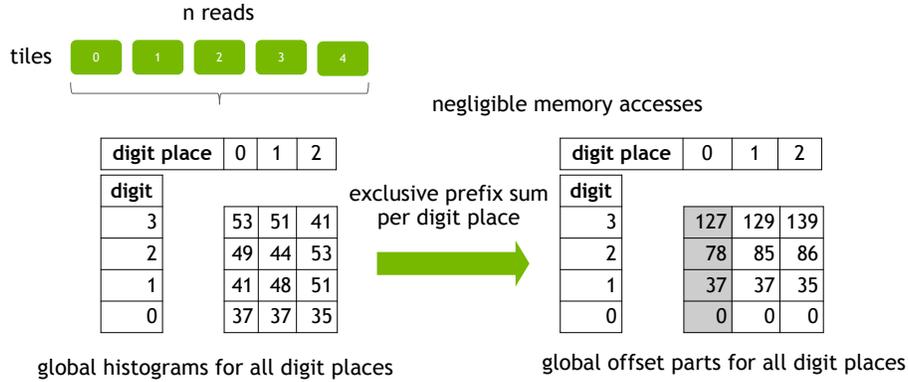

Figure 5. In *Onesweep*, the combined histogram kernel computes digit counts for all digit places for the whole array. Exclusive prefix sum is then applied individually to the histogram for each digit place to produce global parts of the digit offsets. As both depend only on the multi-set of the array elements, which remains unchanged during sorting, they can be computed in a single pass at the beginning, at the cost of n memory reads.

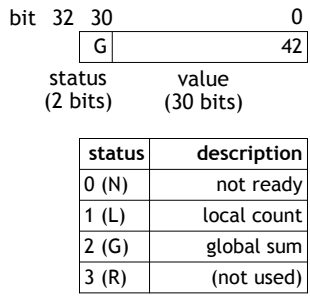

Figure 6. Layout of a single 32-bit counter used in the decoupled look-back implementation of chained scan for the *Onesweep* sort. Two upper bits are used for the status, lower 30 bits contain the value.

*Upfront histogram kernel.* This kernel computes global digit histograms for all digit places. As illustrated in Figure 5, we dispatch a number of thread blocks proportional to the number of SM cores. The input data is tiled among the fixed number of blocks in tile-cyclic fashion. The tile size is the product of two compile-time constants: the number of threads per block, and the number of items per thread. When sorting 32-bit elements on NVIDIA Ampere GPUs, for example, we dispatch blocks of 128 or 256 threads, with each thread consuming 16 or 8 items per tile.

As they consume tiles, the thread blocks within our histogram kernel implementation use atomic addition to compute intermediate, private histograms in shared memory. After consuming their tiles, they atomically accumulate these private histograms into the global histograms in global memory.

We remark that prior implementations of LSD radix sorting upsweep kernels in CUB did not make use of atomic aggregation in shared memory. Histograms were privatized per-thread prior to global aggregation, which required substantial register and shared memory overhead. In contrast, our atomically updated, privatized block-wide histograms use less shared memory. This savings affords bigger histograms, allowing us to increase the radix digit size from the $d = 7$ bits used in CUB v1.10 (128 bins) to $d = 8$ bits (256 bins). This reduces the number of digit places and binning iterations required for *Onesweep* sorting. E.g., for 32-bit keys, the number of binning iterations goes down from 5 to 4. While this increases the cost of each binning iteration, this is more than offset by reducing the number of binning iterations by one.

*Exclusive sum kernel.* This kernel computes a separate exclusive prefix sum across each of the global digit histograms, one per binning iteration. As illustrated in Figure 5, we dispatch one thread block per histogram. As each prefix sum has linear work complexity, the overall work performed by this kernel is proportional to the product of $p$ (digit places) and $r$ (digits). In practice, this kernel consumes an insignificant portion of the total sorting run time, as $p \times r$ is typically $\ll n$.

*Chained scan digit-binning kernel.* This kernel is structured as a variation of the *chained scan* with decoupled look-back technique from Merrill et al. [13]. We dispatch a grid of thread blocks across a tiling of the input, with one block per tile. The tile size is the product of the number of threads per block and the number of items per thread. When sorting 32-bit elements on NVIDIA Ampere GPUs, for example, we typically use 256 or 384 threads per block, and anywhere from 21 to 46 elements per thread.

We use an atomic counter in global memory to assign tiles to thread blocks. This ensures that tiles are processed in order, regardless of the order in which the blocks themselves are scheduled by the underlying hardware.

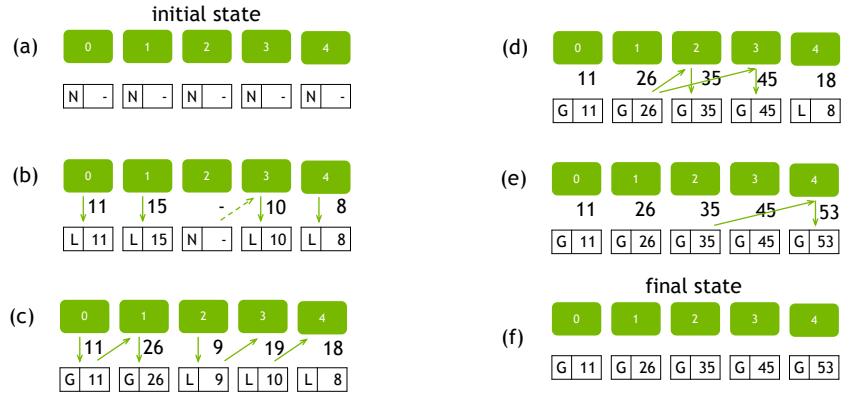

**Figure 7.** An illustration of chained scan with decoupled look-back for *g*=5 tiles and a single digit. (a) In the initial state, all counters contain no value. (b) Except for #2, all counters contain the local count; tile #3 is blocked trying to read the counter #2, which doesn't contain a value. (c) Counters #0 and #1 contain their respective global sums. Tiles #3 and #4 added the local counts of the respective preceding tiles, but don't yet have a global sum. (d) Tiles #2 and #3 now wrote their global sums into their counters. (e) Tile #4 now wrote its global sum to its counter. (f) In the final state, all counters contain the (global) inclusive prefix sum.

The elements of each tile are evenly distributed across the warps of the thread block (where each warp comprises 32 threads). Warps proceed by loading their input elements from global memory and, in the case of signed or floating-point key types, performing a magnitude-preserving conversion to unsigned integer values.

Threads then perform warp-wide key ranking using the *warp-level multi-split* (WLMS) technique described by Ashkiani et al. [2]. For each warp, WLMS computes both (1) a warp-wide histogram of digit-counts, and (2) the warp-relative digit prefix count for each key. We use WLMS because warp-level privatization requires substantially less register and shared memory storage than thread-level privatization, allowing it to scale to higher radixes.

To briefly summarize WLMS, warp threads collectively iterate though their keys in 32-wide batches, with each thread inspecting one key per batch. For each batch, the warp executes a sequence of warp-wide voting operations. For each of the $d$ bits in the current radix digit, the threads within the warp conduct a vote whether that bit is set in their key. By bitwise ANDing the masks it receives for each vote in the current batch, each thread is able to determine the population of warp threads having the same digit as that thread. To accumulate that batch's per-digit-totals in shared memory, the lowest ranked thread in each active digit population will then add the population's count to the per-warp counter corresponding to that digit. That thread then shares the previous count with the other threads in that warp's digit population. Threads can then compute the warp-wide digit prefix count for their key in that batch by adding this batch-wide digit-prefix to the count of lesser-ranked threads in their digit population.

The block then computes an exclusive prefix sum across the per-warp digit counts. This provides (1) tile-wide digit counts for participating in *chained scan* cooperation with other blocks, and (2) tile-relative bin offsets for each warp to scatter its elements into local bins within shared memory. This local reorganization facilitates write-coalescing when subsequently scattering elements to global bins. If necessary, keys are transformed back into signed or floating-point types.

At this point, each block must collaborate with other blocks determine its base scatter offsets. The *chained scan* communication is through a 2D structure of *status counter* arrays, with one array per radix digit, and each array containing one status counter per tile. Thus, each thread block "owns" one status counter for each of the $r$ digits. Although this structure is stored in global memory, it is typically resident in the GPU's L2 cache due to the frequency of access.

Each counter combines two fields within a single 32-bit word: *status* and *value*. This enables coherent, tear-free inspection and modification using simple 32-bit memory operations. The layout of a single counter is depicted in Figure 6. We use the two higher-order bits to store *status*, with the remaining 30 bits storing *value*. The *status* can be 0 (N) if *value* is not ready; 1 (L) if *value* is the local, per-tile digit count; or 2 (G) if *value* is the inclusive global prefix sum of the digit counts of this and all previous tiles.

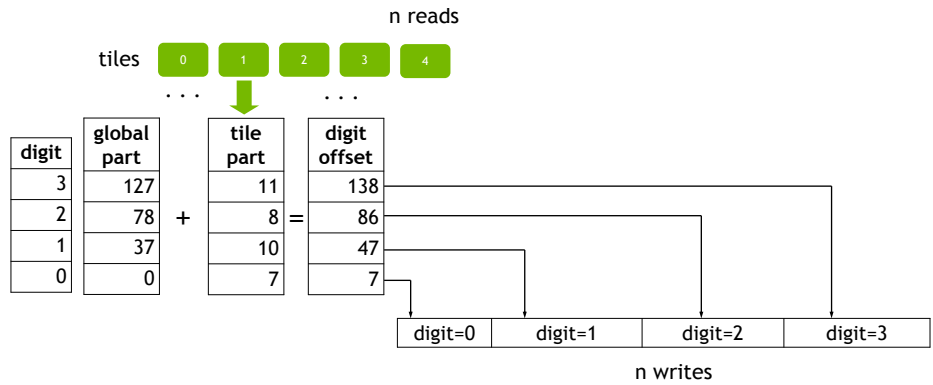

**Figure 8.** A single binning iteration (or partition) of the *Onesweep* sort. The global parts of the tile digit offset have been computed in the combined histogram and exclusive sum kernels. In the partition kernel, chained scans, one per digit, are used to compute the tile part of the offset. When added together (shown here for tile #1 and 4 digits), they give the tile digit offset, at which the thread block processing the tile starts writing its elements with a given digit. Each element is read and written only once, so the partition is performed in 2*n* memory operations (*n* reads and *n* writes).

Figure 7 illustrates a hypothetical timeline of five blocks performing *chained scan* over a single radix digit. Figure 7a shows the initial status counter array in which each counter has *status* 0 (N), indicating its *value* has not yet been updated. In parallel, each block then proceeds to compute its tile-wide digit count. In the counter state shown in Figure 7b, all blocks except #2 have written their tile-wide counts to *value* and updated their *status* accordingly to (L).

After updating its counter with its tile-wide digit count, each block proceeds to compute its exclusive global prefix digit count by reading and adding up the digit counts shared for progressively earlier tiles. If it encounters a tile with a not-ready (N) *status*, it waits for that counter status to change. For example, Figure 7b indicates block #3 must wait on block #2 for the latter to share its tile-wide digit count.

Each block continues this backwards accumulation until it either reaches the beginning of the counter array or until it reads a predecessor counter with *status* (G) whose *value* represents that tile's inclusive global prefix digit count. In the counter state of Figure 7c, for example, block #2 can stop accumulating when it encounters tile #1. At this point, block #2's can calculate its own exclusive global digit prefix count. It then adds its tile-wide count to the exclusive prefix to produce the inclusive prefix, which is then shared through its counter's *value* along with updated *status* (G). For example, blocks #0 and #1 have updated their counter *values* to their inclusive prefix digit counts 11 and 26, respectively.

In the counter state shown in Figure 7d, blocks #2 and #3 have read the inclusive global prefix digit count from tile #1 and updated their counter *values* with inclusive global prefix digit counts 35 and 45, respectively. In Figure 7e, block #4 can read the inclusive global prefix count 26 from tile #2, compute its own global exclusive and inclusive prefixes, and update its counter *value* to 53. At this point, the prefix sum computation is finished; Figure 7f shows the final state with all *values* containing their respective inclusive global prefix digit counts.

In practice, each binning iteration performs $r = 2^d$ concurrent *chained scan* prefix sums, and the $i^{th}$ thread in each thread block is tasked with participating in the chained lookback for digit $i$.

Once the exclusive global digit-relative prefixes are known for a given tile, the block can add them to the global bin offsets to obtain the tile's global bin scatter offsets. Finally, runs of keys are copied from their shared memory bins to global memory bins. If values are sorted alongside the keys, the values are loaded and then scattered using the ranks and digit values computed for the keys.

As an optimization, we can perform a short-circuit check prior to local reorganization. If all keys of the tile have the same digit in the current digit place, we can write them directly to the corresponding output bin using the exclusive global prefix count obtained from *chained scan*.

**Handling very large problems.** In the limit, all *n* keys may end up being distributed into the same digit bin. Consequently, the word sizes of our bookkeeping counters establish an upper bound on the number of items we can sort without incurring counter overflow. In particular, the upfront histogram kernel uses 32-bit counters in shared memory, and the digit-binning kernel uses 30-bit counters in its *chained scan* status arrays. Thus, our base *Onesweep*

**Table 1. Shannon entropy per bit of key data for different entropy bands.**

| # Samples ($q$) | Entropy / bit |
|---|---|
| 1 | 1 |
| 2 | 0.811278 |
| 3 | 0.543564 |
| 4 | 0.33729 |
| 8 | 0.036875 |
| 16 | 0.000266 |

**Table 2. Runtime overheads of *Onesweep* histogram and partition passes for varying digit widths**

| #Partition Passes ($p$) | Bits / Digit place ($d$) | Histogram (µs) | Partition (µs) | Partition / Bit (µs) |
|---|---|---|---|---|
| 6 | 5 | 652 | 1573 | 315 |
| 5 | 6 | 653 | 1685 | 281 |
| 4 | 7 | 679 | 1812 | 259 |
| 4 | 8 | 747 | 1995 | 249 |
| 3 | 9 | 655 | 2541 | 282 |

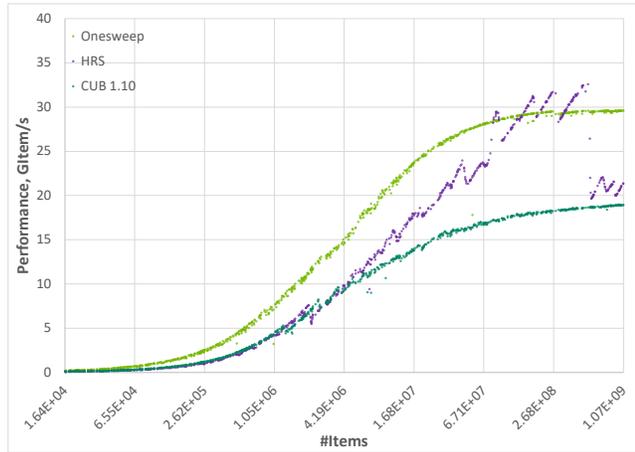

(a) uniform-random keys ($q = 1$)

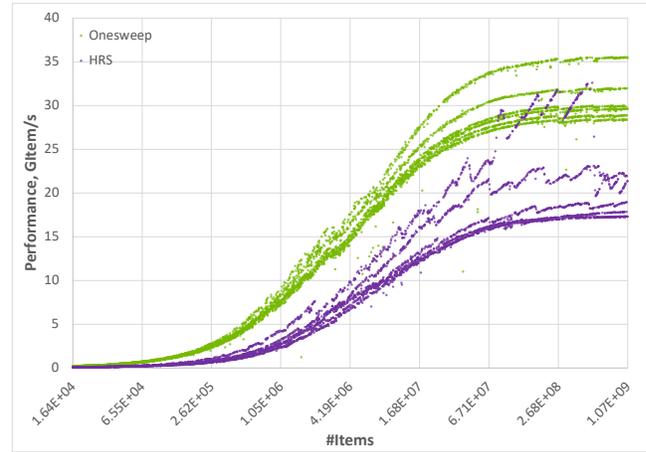

(b) composition of entropy bands

**Figure 9. 32-bit keys-only sorting throughput (NVIDIA A100)**

implementation can only sort problems having less than $2^{30}$ elements.

Unfortunately, the indiscriminate application of wider counter sizes can be a performance impediment, as (1) instruction throughput is lower for 64-bit integer math, and (2) the increased register and shared memory storage overhead would impose smaller, less efficient tile sizes. For the upfront histogram kernel to handle larger problem sizes, we switch to maintaining 64-bit histogram counts in global memory. We continue to use 32-bit counters within the shared memory of each thread block, but instead schedule these blocks to consume their share of the tiles in portions of $2^{30}$ elements or smaller. The 64-bit histograms in global memory are updated after consuming each portion, at which point the block resets its local 32-bit counters.

For the digit-binning kernel to handle large problems, we logically split the input into consecutive strips of $2^{28}$ elements or smaller and then make separate kernel invocations for each strip. The thread block assigned to process the last tile of each strip writes the digit prefixes it had accumulated into a separate global array of 64-bit global bin starting offsets. The thread blocks of the next strip then use those values as the global offset component when computing global scatter offsets. Thus, we can continue to use efficient 32-bit counters for most of our bookkeeping.

## 4    PERFORMANCE EVALUATION

We have implemented *Onesweep* as a replacement LSD radix sorting method within the CUB library of parallel GPU primitives. We evaluate its performance and compare it to the previous CUB v1.10 LSD radix sorting implementation [12] and the HRS MSD implementation [18].

Our test GPU is an NVIDIA Ampere A100-80GB-SXM (A100) with application clocks locked to 1410MHz to prevent dynamic frequency scaling from interfering with performance measurements. We build and run all three implementations (*Onesweep*, CUB, and HRS) specifically for the A100 using the CUDA Toolkit v11.4.1. All tests are performed with ECC enabled.

We use the key-randomization techniques proposed by Thearling and Smith [19] to evaluate the sensitivity of sorting performance to different distributions of key data. Each key is produced by applying bitwise-AND operations across a set of $q$ uniform-random integers. The bit-entropy

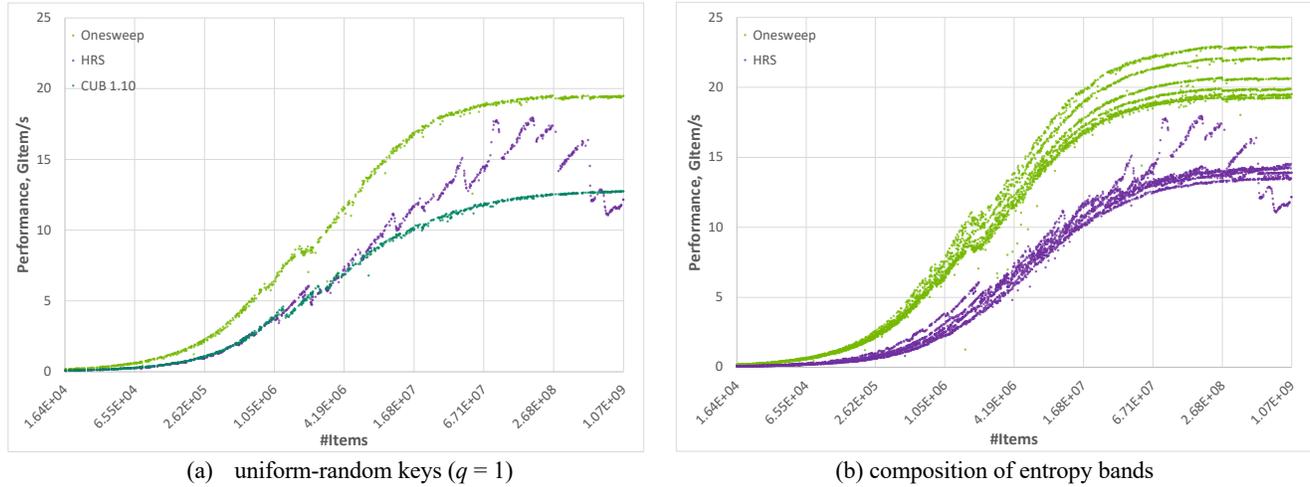

(a) uniform-random keys ($q = 1$)    (b) composition of entropy bands

**Figure 10. 32-bit keys + 32-bit values sorting throughput (NVIDIA A100)**

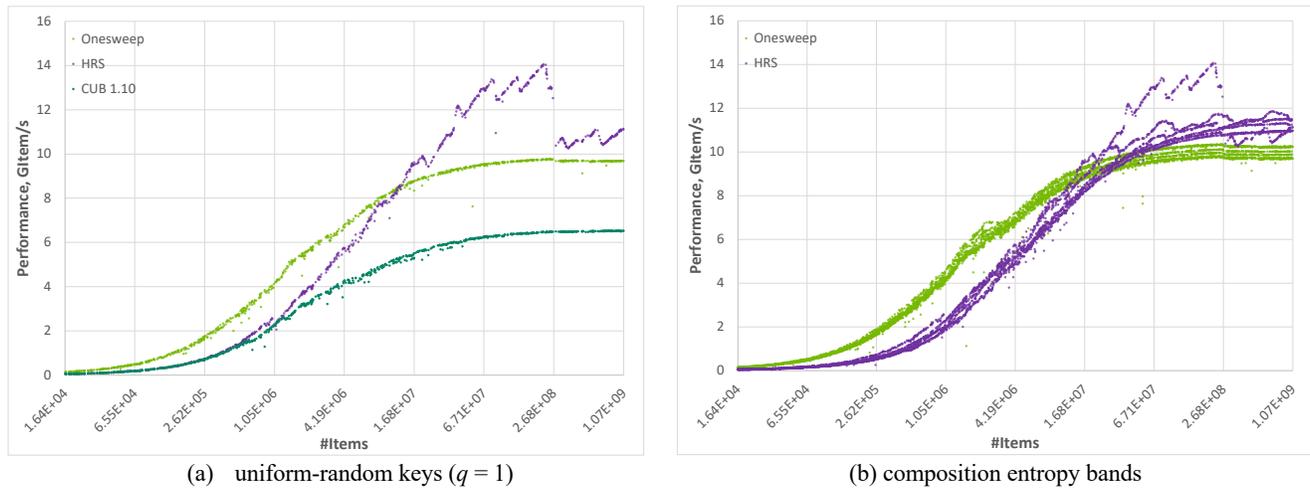

(a) uniform-random keys ($q = 1$)    (b) composition entropy bands

**Figure 11. 64-bit keys-only sorting throughput (NVIDIA A100)**

of the sorting keys is reduced as $q$ increases, leading to binning distributions that are increasingly biased towards digit bins having more 0s in their bit representations. We evaluate problem instances in six different entropy bands, specifically $q = \{1, 2, 3, 4, 8, 16\}$, where $q$ is the number of random samples applied per key. Table 1 enumerates the Shannon entropy per key bit for each band.

We perform ascending-order sorting on 32-bit and 64-bit unsigned integer keys residing in GPU device memory. We also evaluate 32-bit keys paired with 32-bit payload values. Problem sizes are randomly sampled on a log-scale from $2^{12}$ to $2^{30}$. For algorithm, key-value type combination and entropy-band, we sample 1000 different problem sizes.

***Performance for different problem sizes and distributions.*** Figure 9, Figure 10, and Figure 11 plot sorting throughput as a function of problem size. As we would expect from the linear work complexity of radix sorting, the performance of the *Onesweep* and CUB LSD implementations exhibit a smoothed roofline trend in which throughput increases with problem size, plateauing when computational and memory I/O resources become saturated. The performance response of HRS is "choppier" due to the local-finishing aspect of the MSD method, which provides outsize benefits for certain small problem size regimes, especially those with uniformly random key distributions.

*Onesweep* is unilaterally faster than the LSD radix sort from CUB 1.10. For uniform-random key distributions, we demonstrate a speedup range of 1.4-1.6x for saturating problem sizes. This compares favorably with the theoretical *Onesweep* speedup ceiling, which for 32-bit keys is 1.67x due to (1) the elimination of digit histogram passes and (2) one less digit binning iteration.

Onesweep also provides a clear performance advantage over the HRS implementation of MSD radix sorting for 32-bit keys. As illustrated in Figure 10, *Onesweep* is always better when paired with values. Notably, Figure 9a highlights a few regimes where HRS is marginally faster when sorting uniformly distributed 32-bit keys-only. However, HRS performance drops considerably for problems exceeding 550M items, which impose an additional round of global binning before local finishing. Furthermore, Figure 9b reveals that even a small amount of distribution bias ($q > 1$) will delay the onset of widespread local finishing in HRS, resulting in its substantially lower performance.

*Onesweep*, on the other hand, demonstrates a much greater degree of performance consistency when subjected to increased key entropy. Figure 9b shows that, for 32-bit keys-only sorting, Onesweep sorting throughput drops 5% from 29.4 GKey/s for uniform-random keys ($q = 1$) to 28 GKey/s for highly skewed keys ($q = 8$). The performance drop for 64-bit items is even smaller. When entropy decreases to the point where most tiles are homogenous ($q = 16$), our copy-optimization demonstrates throughput as high as 36 GKey/s for 32-bit keys-only sorting.

For 64-bit keys, HRS has a clear advantage for large, saturating problem sizes. Whereas *Onesweep* performs eight binning iterations in global memory, HRS only requires four or five iterations before sub-problems are small enough for local finishing. The performance gap between *Onesweep* and HRS is much smaller in the presence of distribution bias ($q > 1$). For problems too small to saturate the GPU, the larger bookkeeping overhead of MSD radix sorting is an outsize cost for HRS.

***Cost of individual passes.*** Table 2 provides a runtime breakdown of *Onesweep* kernels. We list runtimes for the upfront histogram and digit-binning kernels when sorting 256M random 32-bit keys using a range of different digit widths *d*. We observe that the upfront histogram kernel is relatively insensitive to *d*. Furthermore, as it is invoked only once per sort, its contribution to the overall sorting time is relatively small. The runtime of digit-binning, on the other hand, increases significantly with the digit size. When normalizing by the number or bits processed per pass, we find the greatest efficiency (in terms of per-bit partitioning time) is achieved for 8-bit partitions.

## 5   CONCLUSION

In this paper, we presented *Onesweep*, an improvement to the state-of-the-art of GPU LSD radix sort. By using a single pass of digit-binning based on *chained scan* with decoupled look-back, we reduce the number of memory operations from *3n* to *2n* per binning iteration. We also reduce the number of binning iterations by using a larger digit size.

This provides a consistent improvement of 1.4x-1.6x versus CUB version 1.10, the current state-of-the-art in GPU LSD radix sorting. On the NVIDIA A100 GPU, it achieves sorting speeds of 29.4 GKey/s for uniform-random 32-bit keys. Furthermore, *Onesweep* provides substantially greater performance consistency than HRS, the state of the art in GPU MSD radix sorting, while outperforming it in almost all cases.

MSD's local-finish capability has the alluring appeal of fast performance, especially on smaller problems with larger key data types. In practice, however, the challenges of achieving high levels of utilization from irregular workloads yields a very inconsistent performance response that is highly dependent upon the key distribution and problem size. Conceptually, it should outpace LSD for all small problems, but we show that the overheads of sub-problem management prevent it from doing so. LSD on the other hand, provides a much more consistent performance response, which is desirable when modeling application workloads, especially in embedded and/or real-time contexts.